\documentclass[aps,a4paper,showpacs,twocolumn,superscriptaddress]{revtex4}
\usepackage{graphicx}
\usepackage{amsmath}
\usepackage{amssymb}
\usepackage{enumerate, color}
\usepackage{subfigure}
\usepackage{tabularx}
\usepackage{epstopdf}
\usepackage{lipsum}
\usepackage{slashed}

\newcommand{\be}{\begin{equation}}
\newcommand{\ee}{\end{equation}}
\newcommand{\ben}{\begin{eqnarray}}
\newcommand{\een}{\end{eqnarray}}
\newcommand{\bes}{\begin{subequations}}
\newcommand{\ees}{\end{subequations}}

\newcommand{\sech}{{\rm sech}}

\begin{document}
\title{Fermion bound states in geometrically deformed backgrounds}
\author{D. Bazeia}
\affiliation{Departamento de F\'isica, Universidade Federal da Para\'iba, 58051-970, Jo\~ao Pessoa, PB, Brazil}
\author{A. Mohammadi}
\affiliation{Departamento de F\'isica, Universidade Federal de Pernambuco, 52171-900, Recife, PE, Brazil}
\author{D. C. Moreira}
\affiliation{Departamento de F\'isica, Universidade Federal da Para\'iba, 58051-970, Jo\~ao Pessoa, PB, Brazil}

\begin{abstract}
This work deals with the behavior of fermions in the background of kinklike structures in the two-dimensional spacetime. The kinklike structures appear from bosonic scalar field models that engender distinct profiles and interact with the fermion fields via the standard Yukawa coupling. We first consider two models that engender parity symmetry, one leading to the exclusion of fermion bound states, and the other to the inclusion of bound states, when the parameter that controls the bosonic structure varies from zero to unity. We then go on and investigate another model where the kinklike solution explicitly breaks parity symmetry, leading to fermion bound states that are spatially asymmetric.      
\end{abstract}

\maketitle

\section{Introduction}

The study of fermions in the presence of kinklike structures have been initiated long ago, in the pioneering work by Jackiw and Rebbi \cite{jr}. An important information that appears from the investigation is the phenomenon known as fermion number fractionalization, which is due to the topological nature of the background bosonic structure \cite{gol}. The model investigated in \cite{jr} is defined in $(1,1)$ spacetime dimensions, and describes a real scalar field that interacts with a fermion field via the Yukawa coupling. For more on kinks and related issues see, e.g., Ref.~\cite{vachaspati}.

The interest in the fermion number fractionalization goes beyond its mathematical identification since it presents peculiarities that can be physically realized in condensed matter situations, as shown in Refs.~\cite{ssh,js}. The subject has been investigated by other authors, and here we quote Refs.~\cite{shifman,brihaye,charmchi,AA,bmohammadi} to illustrate this possibility. As one knows, however, the effect of the fermion number fractionalization is directly related to the topological behavior of the bosonic structure arising from the bosonic portion of the model. However, in the recent work \cite{bmohammadi} one investigates another possibility, focusing attention on the geometric conformation of the topological structure that the bosonic portion of the model brings into play.

The geometrical aspects of the structure is of current interest, since experiments may now be carried out on miniaturized samples in constrained geometries, and the geometry may drastically change the conformational structure of the topological object, as experimentally verified for instance in Ref.~\cite{cg}. The change in the conformational structure of the bosonic background may induce distinct physical properties on the fermion field, as it was presented in \cite{bmohammadi} and is further shown in the current work.

There are many motivations to study the interaction of fermion fields with bosonic backgrounds since it may create or affect other interesting physical phenomena like the Casimir effect \cite{ce1,ce2}, the Bose-Einstein condensation \cite{be}, and the localization of fermions in braneworld scenarios \cite{bs1,bs2,bs3,bs4}. Another motivation is the current interest in the study of miniaturized samples of magnetic materials \cite{cg,NN,Sci,nano,PRA} and the recent investigation \cite{bmohammadi}. With this in mind, here we introduce three models of the type considered in \cite{jr} which support distinct bosonic backgrounds. In the models, the bosonic portion that generates the topological structures was studied before in Refs.~\cite{bm,bmvaclss,bmmashyb}, and we use them to describe how the fermion field behaves in such distinct backgrounds. 

To implement the investigation, we organize the work as follows: In Sec.~\ref{gen} we introduce the general model and deal with some of its properties, of direct interest to the current investigation. We move on and review the case of a fermion field coupled to the sine-Gordon model in Sec.~\ref{models}, since this is also of general interest to the current work. Then, we study the three new models, explicitly showing how the fermion bound states and energies are characterized in each case. In the two first models, the bosonic background structures are controlled by a real parameter and obey the parity symmetry, but they behave differently as the parameter increases from zero to unity, one excluding and the other including fermionic bound states in the system. The third model is different, and the bosonic structure does not obey the parity symmetry anymore, so all the fermion bound states are asymmetric functions. We end the work in Sec.~\ref{end}, where we add some comments and conclusions.

\section{Generalities}
\label{gen} 

We are interested in studying models described by the Lagrangian
\begin{equation}\label{L}
\mathcal{L}=\frac12\partial_\mu\phi\partial^\mu\phi-V(\phi)+\frac{1}{2}\bar{\psi}i\slashed{\partial}\psi-\phi\bar{\psi}\psi,
\end{equation}
which is similar to the model of Ref.~\cite{jr}. We are dealing with a scalar field represented by $\phi=\phi(x,t)$ and a Dirac field denoted by
$\psi=\psi(x,t)$, which interact via the Yukawa coupling that appears in the last term of the above expression. In the models to be considered here we define $V(\phi)=W_\phi^2/2$, where $W_\phi$ is the derivative of some function $W=W(\phi)$ with respect to the field $\phi$. $W$ in supersymmetric models is called superpotential, although here we are using it as a mathematical tool to simplify the calculations. Also, we use $\hbar=1=c$ and consider dimensionless fields and the spacetime coordinates.

In this sense, we consider the topological structure of the kinklike profile which arises from the bosonic Lagrangian 
\ben\label{bm}
\mathcal{L}_b&=&\frac{1}{2}\partial_\mu\phi\partial^\mu\phi-\frac12 W_\phi^2,
\een
as background solutions to be considered in the fermionic Lagrangian
\ben \label{lagf}
\mathcal{L}_f&=&\frac{1}{2}i\bar{\psi}\slashed{\partial}\psi-\phi\bar{\psi}\psi.
\een
The procedure goes as follows: we first deal with the bosonic model \eqref{bm} to find the static kinklike structure that solve the corresponding equation of motion
\be
\phi''-W_\phi W_{\phi\phi}=0,
\ee
where the prime stands for derivative with respect to the spatial coordinate $x$. As it is well-known, in this case the solutions obey the first-order equation 
\be 
\phi^\prime=W_\phi,
\ee
and so are stable against small fluctuations. 

The equations of motion for the fermion field has the form
\be\label{eom}
\left(i\slashed{\partial}-2\phi\right)\psi=0.
\ee
For convenience, we choose to describe the gamma matrices by the set $\left(\gamma_0,\gamma_1,\gamma_5\right)=\left(\sigma_1, i\sigma_3,\sigma_2\right)$. Moreover, since the scalar field describes static structure we write the spinor field as
$$\psi(x,t)=e^{-iEt}\left(
\begin{array}{c}
\psi^{(+)}(x)\\
\psi^{(-)}(x)\\
\end{array}\right).$$
This ansatz can be inserted in Eq.~(\ref{eom}) which allows us to rewrite the equation of motion for the Dirac field in the form
\begin{equation}\label{coupledsystem}
E\psi^{(\pm)}+\left(\pm\frac{d}{dx}-2\phi\right)\psi^{(\mp)}=0,
\end{equation}
which is a system of equations involving the components of the spinor field $\psi$.
We can use this system of equations to obtain two Schr\"odinger-like equations given by
\begin{equation}\label{decoupledsystem}
\left(-\frac{d^2}{dx^2}+U_{(\mp)}(x)\right)\psi^{(\pm)}=E^2\psi^{(\pm)},
\end{equation}
where $U_{\pm}(x)=\pm 2d\phi/dx+4\phi^2$, with $\phi=\phi(x)$ being the static kinklike solution of the bosonic system. These decoupled Eqs.~\eqref{decoupledsystem} and the first order Eqs.~{\eqref{coupledsystem}} are used to find the bound states and energy spectrum of the fermion system.

We note that equations (\ref{decoupledsystem}) have the form $Q^{\mp}Q^{\pm}\psi^{(\pm)}=E^2\psi^{(\pm)}$, where $Q^{\pm}=\pm d/dx+2\phi$. In particular, we can find an expression for the ground state wavefunction when solving $Q^{\pm}\psi^{(\pm)}=0$, obtaining the result
\begin{equation}
\psi^{(\pm)}_0=c_{\pm} e^{\mp2\int^x \phi(x')dx'},
\end{equation}
where $c_{\pm}$ are normalization constants, and for regularity of the ground state one of them has to be zero. To find the massive bound states, one uses Eqs.~\eqref{coupledsystem} and \eqref{decoupledsystem}. It can also be shown that the stability equation for the scalar field is
\begin{equation}\label{zero-mode}
\left(-\frac{d^2}{dx^2}+U(x)\right)\eta_n(x)=\omega^2_n\eta_n(x),
\end{equation}
where
\be 
U(x)=\left.\frac{d^2 V}{d\phi^2}\right|_{\phi=\phi(x)}.
\ee
To get to the above equation, we have set $\phi(x,t)=\phi(x)+\sum_n\eta_n(x)\cos(\omega_n t)$. In this case the zero mode $\eta_0(x)$ of the scalar field is proportional to the derivative of the static solution itself, i.e, $\eta_0(x)\approx\phi'(x)$, and the solution is classically or linearly stable. To see this explicitly, one recalls that 
\be
U(x)=\frac{d^2 V}{d\phi^2}=W_{\phi\phi}+W_\phi W_{\phi\phi\phi},
\ee
which has to be calculated at the classical static solution $\phi=\phi(x)$. Thus, one can rewrite the second-order differential operator in Eq.~\eqref{zero-mode} as
\be 
-\frac{d^2}{dx^2}+ W_{\phi\phi}W_{\phi\phi\phi}=\left(-\frac{d}{dx}-W_{\phi\phi}\right)\left(\frac{d}{dx}-W_{\phi\phi}\right)
\ee
so it is a non-negative operator.

Another interesting result is that the threshold energy is taken at the limit $\phi(x\to\infty)=\phi_{min}$, where the bosonic field approaches the minimum of the scalar potential. Moreover, for the well-behaved solutions of the fermionic bound states we require that in this regime $\psi^{(\pm)}\to c_{\pm}$ and $d\psi^{(\pm)}/dx\to 0$. In this way, the threshold energy equation, derived from (\ref{coupledsystem}), becomes $E_{th}c_{\pm}-2\phi_{min}c_{\mp}=0$, and thus one finds $E_{th}=2\phi_{min}$, which is equal to the square root of the value of $U_{\pm}$ at spatial infinities, as expected. 

\section{Models}
\label{models}

Now, let us consider some explicit models. 
We are interested in studying the fermion field behavior when it evolves in the background of a kinklike structure derived from the bosonic model defined in terms of the potential 
\be  
V(\phi)=\frac12\, W_\phi^2,
\ee 
for the following three distinct cases: 
\bes\label{wmodels}\ben
W_\phi&=&\frac{1}{1-\lambda}\text{cd}(\phi,\lambda),\\
W_\phi&=&\frac{2\text{cn}^2(\phi/2,\lambda)-(1-\lambda)}{\text{dn}(\phi/2,\lambda)},
\een\ees
and
\be\label{asy}
W_\phi=(1-\phi)(1+\phi^p).
\ee

Kinklike solutions for the bosonic models (\ref{wmodels}a) and (\ref{wmodels}b) were presented in \cite{bm,bmvaclss}. The models are written in terms of Jacobi's elliptic functions, where $\text{cd}(\phi,\lambda)=\text{cn}(\phi,\lambda)/\text{dn}(\phi,\lambda)$ and $\lambda$ is a real parameter in the interval $[0,1]$. Both (\ref{wmodels}a, \ref{wmodels}b) retrieve the sine-Gordon model for $\lambda\to 0$ and approach solutions with infinite amplitude when $\lambda\to 1$, but in quite different ways.
The model (\ref{wmodels}a) has, for any value of $\lambda$, an infinite set of degenerate topological sectors with meson mass $m^2=1/(1-\lambda)^2$, which increases as $\lambda\to 1$ and is not defined at $\lambda=1$. The model (\ref{wmodels}b) has two different infinite sets of topological sectors, but the mass of the meson is $m^2=4(1-\lambda^2)$, which decreases as $\lambda\to1$  and is well defined at $\lambda = 1$, where it is zero. For $ \lambda=0$ these sets are equivalent, but they are different as we vary the $\lambda$ parameter. In particular, the topological sector we choose to work on here approaches the vacuumless solution in the limit $\lambda\to1$.

The third model is defined by Eq.~\eqref{asy}. It was presented in \cite{bmmashyb}, and the parameter $p$ is an odd integer, $p=1,3,5,...$ . Here the system presents a single topological sector, and the reflection symmetry is broken for $p\neq1$. In this case, we do not have changes in the minima of the scalar potential, which are at $\phi= \pm 1$ for all  $p$, so that the asymmetry is only revealed by the two classical meson masses, or by the potential seen by the fermion field.

Considering the lagrangian (\ref{lagf}), for all three models the system has energy-reflection symmetry given by $\gamma^1$ as well as charge-conjugation symmetry which is representation dependent and in the representation we have chosen is given by $\sigma_3$. Therefore, we expect that the fermionic bound energy spectrum is symmetric around the $E=0$ line in all cases to be considered here. However, althought the first two models enjoy parity or reflection symmetry, the third model does not respect this symmetry, and so it should be studied more carefully.

Due to the relevance of the sine-Gordon model \cite{caudrey} in the context of the current work, let us first review its solution and stability. It appears as a particular case of the models (\ref{wmodels}a) and (\ref{wmodels}b) for $\lambda = 0$. Thus, we have $W_\phi=\cos \left(\phi\right)$ and the solution for the scalar field has the form
\begin{equation}\label{sGsol}
\phi(x)=\pm\sin^{-1} \left(\tanh (x)\right).
\end{equation}
In this case, the stability potential associated with the bosonic field is given by $V_{SG}=1-2 \text{sech}^2(x)$, which has a reflectionless shape and only one bound state, the zero mode, given by $\eta_0=\sech(x)$. However, if one takes the above solution and uses it into the equation \eqref{decoupledsystem}, we end up with the following potentials
\begin{widetext}{\center
\begin{figure}
\centering 
\begin{tabular}{cc}    
\subfigure[$E_1=\pm1.87806$]{\label{fig:a}\includegraphics[width=60mm]{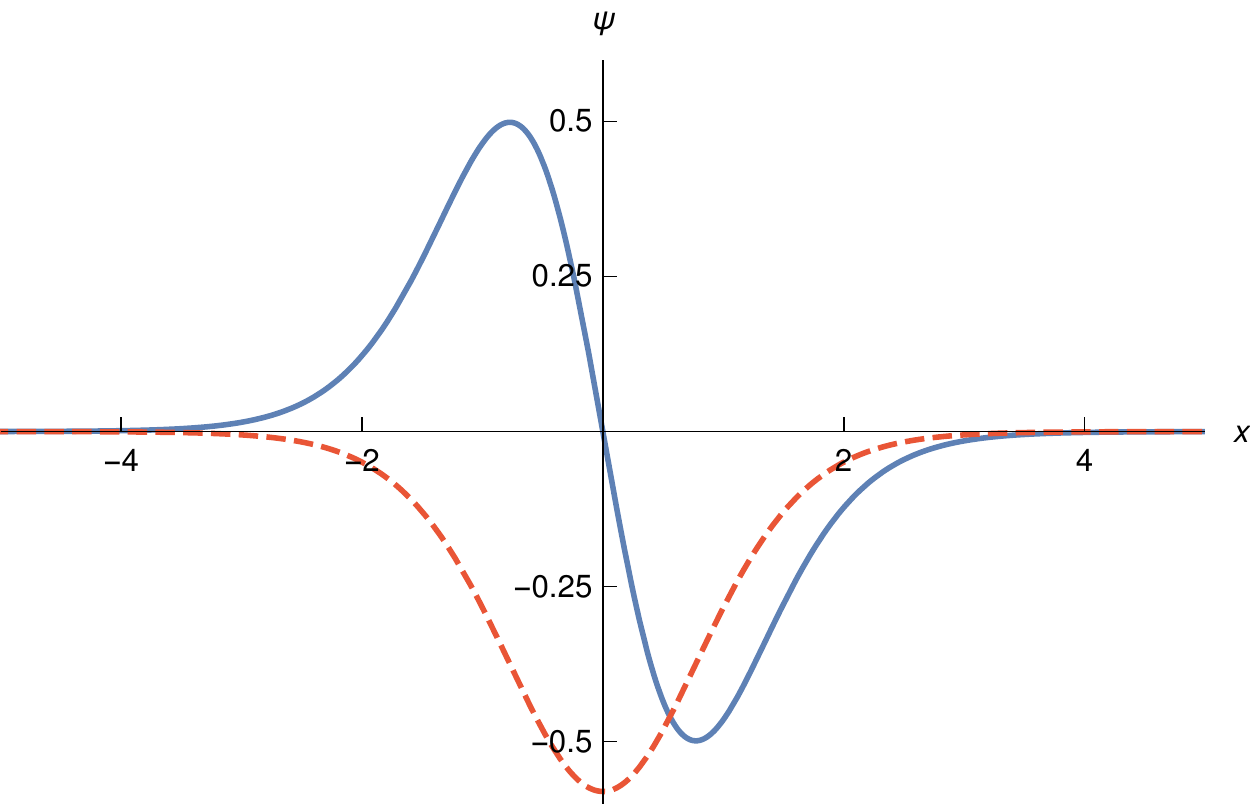}}\;\;\;\;\;
\subfigure[$E_2=\pm2.48335$]{\label{fig:b}\includegraphics[width=60mm]{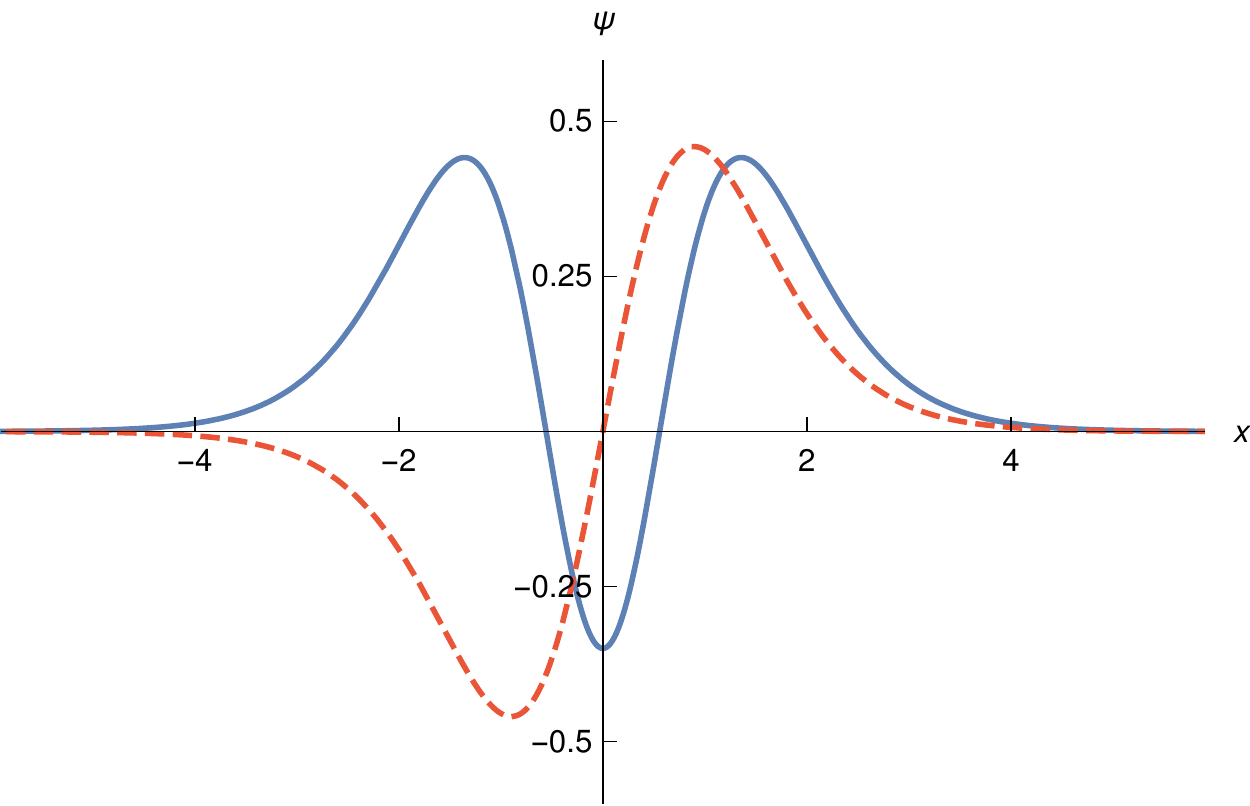}}
\end{tabular}
\begin{tabular}{cc}
\subfigure[$E_3=\pm2.83358$]{\label{fig:b}\includegraphics[width=60mm]{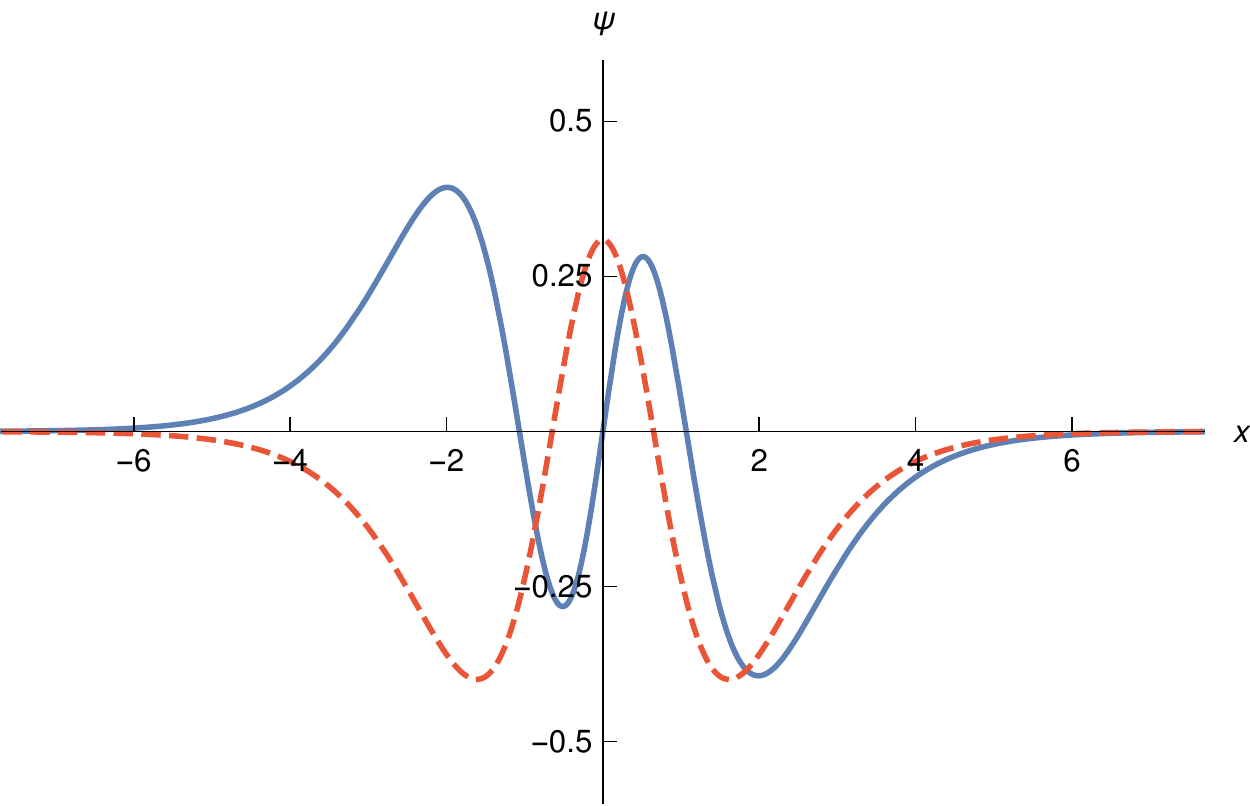}}\;\;\;\;\;
\subfigure[$E_4=\pm3.03448$]{\label{fig:b}\includegraphics[width=60mm]{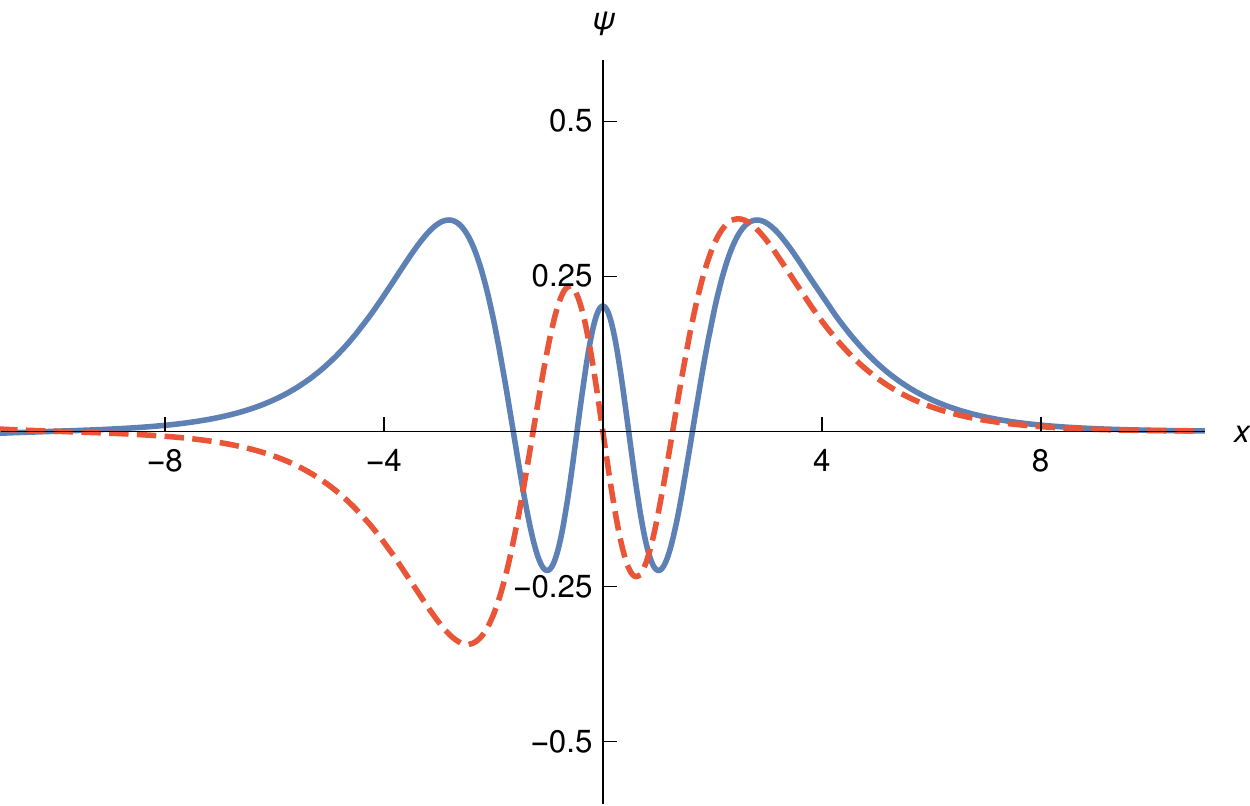}}
\end{tabular}
\caption{The $\psi^{(+)}$ (blue, solid line) and $\psi^{(-)}$ (red, dashed line) components of the massive bound states and the corresponding eigenenergies of the fermion field coupled to the sine-Gordon soliton (\ref{sGsol}).}\label{fig1}
\end{figure}}\end{widetext}

\begin{equation}\label{UpmsG}
U_{\pm}=4 \left(\sin ^{-1}\left(\tanh (x)\right)\right)^2\pm2 \text{sech}(x).
\end{equation}
It asymptotically approaches $U_{\pm}(\pm\infty)=\pi^2$. The potential $U_-(x)$ allows nine fermionic bound states, which occur at the energies $E_0=0, E_1=\pm1.87806, E_2=\pm2.48335, E_3=\pm2.83358$ and $E_4=\pm3.03448$. The zero mode can be obtained analytically and, up to a normalization factor, is given by
$$\psi_0(x,t)\propto\!\!\left(\!
\begin{array}{c}
e^{-\left(2x \left(2 \cot ^{-1}\left(e^x\right)+\sin ^{-1}(\tanh (x))\right)+2\text{Ti}_2\left(e^{-x}\right)\right)}\\
0\\
\end{array}
\!\right).$$
Here $\text{Ti}_2\left(x\right)$ is the inverse tangent integral, which can be written in terms of polylogarithmic functions by the relation $\text{Ti}_2\left(e^{-x}\right)=i \left(\text{Li}_2\left(-i e^{-x}\right)-\text{Li}_2\left(i e^{-x}\right)\right)$. For the other bound states, we solve the set of equations in (\ref{coupledsystem}) and \eqref{decoupledsystem} and plot them in Fig.~\ref{fig1}.

\begin{figure}[t]
\centerline{\includegraphics[height=15em]{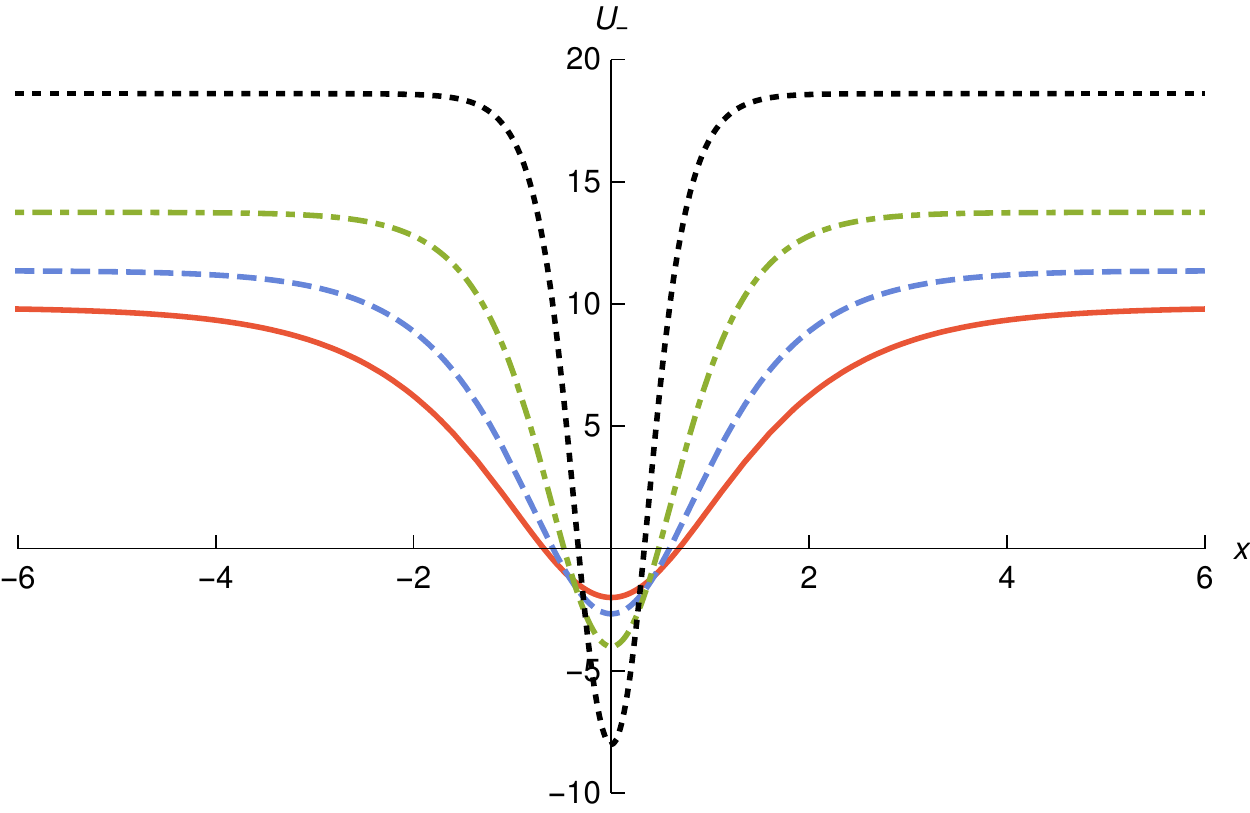}}
\caption{The $U_-$ potential that appears in Eq. \eqref{Upmmodel1} for the model I, depicted for $\lambda=0,0.25,0.5,0.75$ with solid, dashed, dot-dashed and dotted curves respectively.}\label{fig2}
\end{figure}

\subsection{Model I}

Let us now look at the model (\ref{wmodels}a) for general $\lambda$. In this case, the solution obtained for the scalar field is given by
\begin{equation}\label{solmodel1}
\phi(x)=\text{sn}^{-1}\left(\tanh \left(\frac{x}{1-\lambda }\right), \lambda \right)
\end{equation}
where $\text{sn}(x,\lambda)$ is the Jacobi elliptic sine, and its stability potential is
\begin{equation}\label{vbosonm1}
U(x)\!=\!\frac{1 -\left(\!1-\!\frac{1}{1-\lambda}\cosh \!\left(\frac{2 x}{1-\lambda}\right)\right)\!\text{sech}^4\left(\frac{x}{\!1-\lambda}\right)}{\left(1-\lambda\tanh ^2\left(\!\frac{x}{1-\lambda }\!\right)\right)^2}.
\end{equation}
It approaches $U(x\to\pm\infty)\sim 1/(1-\lambda)^2$, which implies that the depth of the well increases as $\lambda$ increases. For  $\lambda=0$ the expression (\ref{vbosonm1}) is well defined and has only one bound state which is the sine-Gordon case. However, for the other values of $\lambda$, we find an excited state, which has an energy gap with respect to the ground state that increases as $\lambda$ increases. 

\begin{figure}[t]
\centerline{\includegraphics[height=14em]{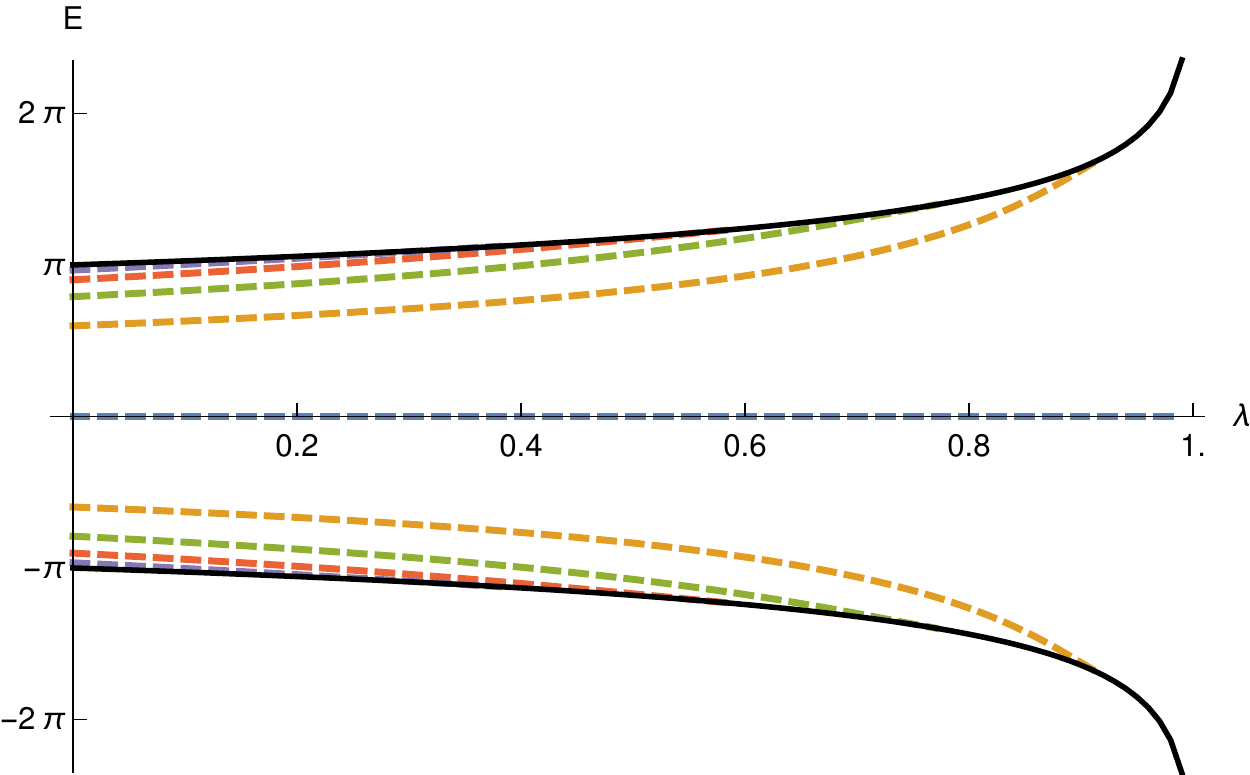}}
\caption{The fermionic bound state energy spectrum as a function of $\lambda$ for the model I. The solid black curves identify the threshold energies.}\label{fig3}
\end{figure}
\begin{figure}[t]
\centerline{\includegraphics[height=15em]{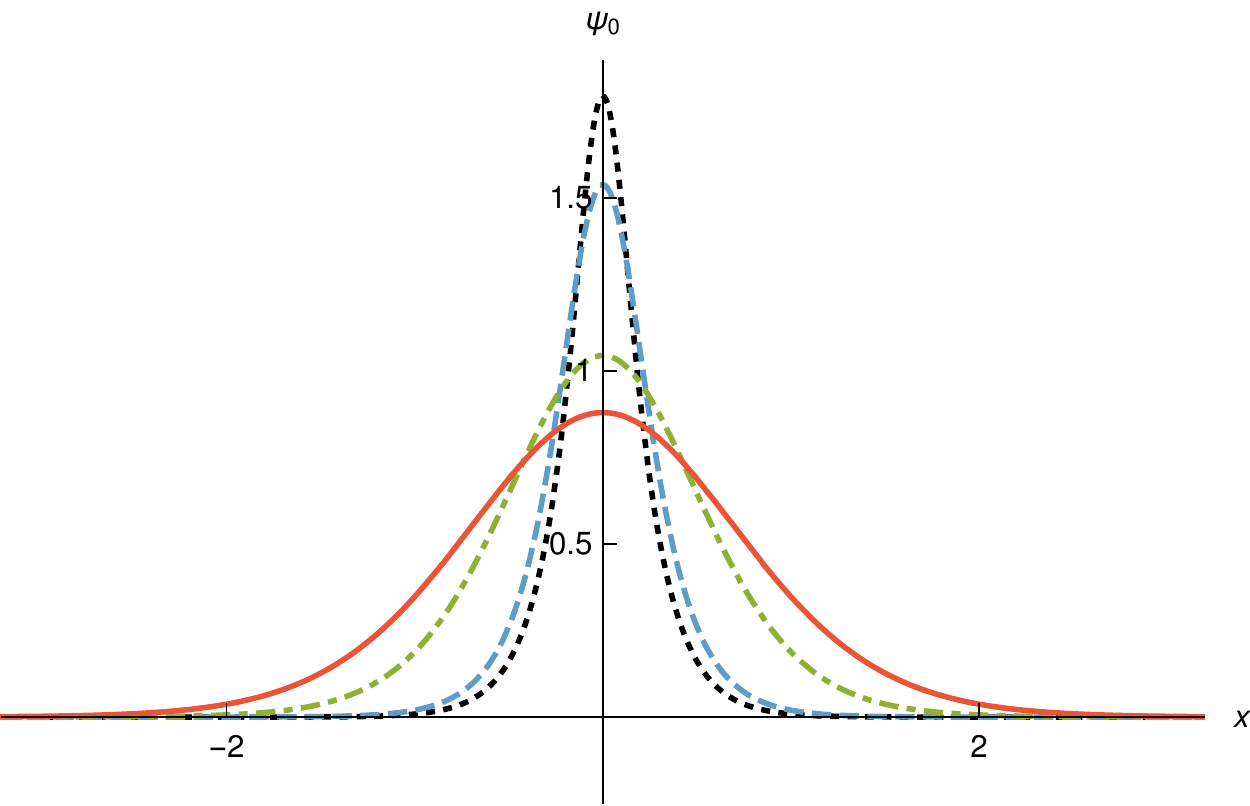}}
\caption{The normalized fermion zero mode derived from equations (\ref{coupledsystem}) and \eqref{decoupledsystem} with scalar field given by \eqref{solmodel1} for $\lambda=0,0.5,0.9,0.95$ with solid, dashed, dot-dashed and dotted curves respectively, for the model I.}\label{fig4}
\end{figure}

Given the Yukawa coupling between boson and fermion fields, the Dirac field spectrum must be affected by changes in the behavior of the bosonic structure. For the model we are now analyzing, the fermionic eigenstates are given by equations \eqref{coupledsystem} and \eqref{decoupledsystem}, and now the potentials $U_{\pm}$ have the forms
\begin{eqnarray}\label{Upmmodel1}
\nonumber U_{\pm}&=&4 \left(\text{sn}^{-1}\! \left(\tanh\!\left(\!\frac{x}{1-\lambda }\!\right)\!\!,\!\lambda\!\right)\right)^2\pm\\
&~&\pm\frac{2}{1-\lambda}\text{cd}\!\left(\!\text{sn}^{-1}\!\left(\!\tanh\!\left(\frac{x}{1-\lambda}\!\right)\!\!,\!\lambda \right)\!,\!\lambda\!\!\right).
\end{eqnarray}
The behavior of $U_-$ is depicted in Fig. \ref{fig2}. Asymptotically, it approaches $U_{\pm}(\pm\infty)=4 K(\lambda )^2$, where $K(\lambda )$ is the complete elliptic integral of the first kind, which diverges as $\lambda\to 1$. At $x=0$ one gets $U_{\pm}(0)=\pm 2/(1-\lambda)$. Therefore, for $U_-$ the depth of the well increases and its width diminishes significantly as $\lambda\to1$. This effect causes the exclusion of bound states in the well, as we illustrate in Fig.~\ref{fig3}. In particular, one can see that for the following values of $\lambda$ there are the respective numbers of bound states: for $\lambda=1/4$, nine bound states; for $\lambda=1/2$, seven bound states; for $\lambda=3/4$, five bound states; and for
$\lambda=9/10$, only one bound state. To find numerically the bound states in this model and the other two as well, we solved the eigenvalue problem of Eq.~\eqref{decoupledsystem} using Mathematica. Besides that, we confirmed the results solving the first order differential equations in \eqref{coupledsystem}, where we have adopted the Runge Kutta Fehlberg order 5 method.

Unfortunately, we can not find the analytical expression for the bound state wave functions corresponding to an arbitrary $\lambda$. Nevertheless, we can observe some characteristics of its behavior. In the vicinity of $x = 0$ the scalar field behaves as $\phi(x\to0)\simeq x/(1-\lambda)+\mathcal{O}\left(x^3\right)$, so the shape of the ground state in this region is $\simeq e^{-x^2/(1-\lambda)+\mathcal{O}\left(x^4\right)}$, which means the higher the value of $\lambda$, the narrower the wave function is. Moreover, asymptotically the scalar field is $\phi(x\to\infty)\simeq e^{-x/(1-\lambda)}+K(\lambda)$, which implies a decay proportional to $e^{-2K(\lambda )x}$ in the ground state wave fucntion as $\lambda\to1$. The behavior of the fermionic zero mode wave function in these two limits suggests that as $\lambda$ increases, the normalized wave function becomes taller and narrower, as illustrated in Fig.~\ref{fig4}. The same effect occurs for the excited states of the model.

\subsection{Model II}
\begin{figure}[t]
\centerline{\includegraphics[height=14em]{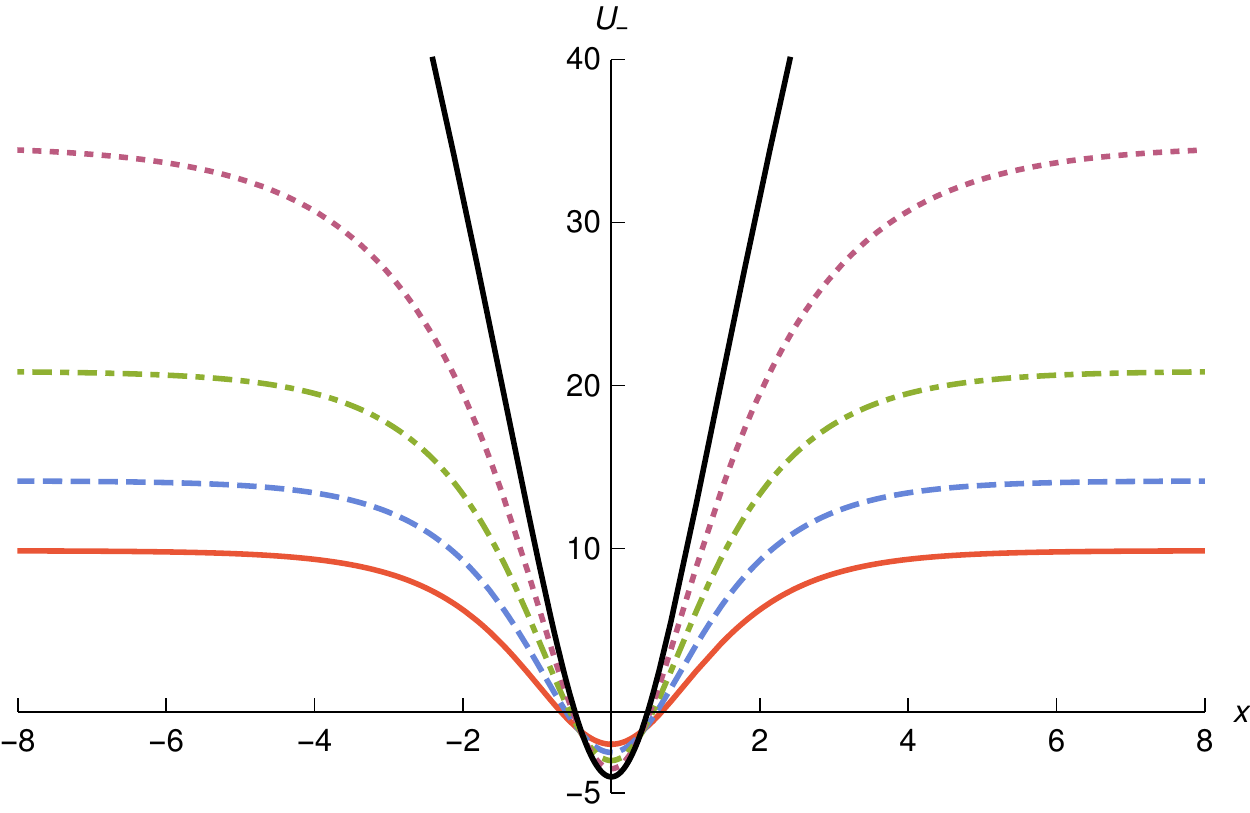}}
\caption{The $U_{-}$ potential that appears in (\ref{Upmmodel2}) for the model II, depicted for $\lambda=0,0.25,0.5,0.75,$ and $1$, with solid (red), dashed, dotdashed, dotted and solid (black) curves respectively.}\label{fig5}
\end{figure}

We now study the model (\ref{wmodels}b) for general $\lambda$. As shown in \cite{bmvaclss}, this model has two solutions, and in one of them there is a transition between sine-Gordon kink and vacuumless solution presented in \cite{vilenkin1,bvacless}, so we choose this solution as background field. In formula, the background field is given by
\begin{equation}\label{solmodel2}
\phi(x)=2 \text{sc}^{-1}\left(\sqrt{\frac{1+\lambda}{1-\lambda}} \tanh \left(\frac{1}{2}\sqrt{1-\lambda ^2} x \right),\lambda \right)
\end{equation}
where $\text{sc}(\phi,\lambda)=\text{sn}(\phi,\lambda)/\text{cn}(\phi,\lambda)$. Here we have $\lambda\in[0,1]$, and the stability potential for the scalar field, which has only one bound state for any $\lambda$, interpolates between a reflectionless shaped potential, for the sine-Gordon case, and a volcano potential for the vacuumless solution. The drastic change in the shape of the stability potential can be explained by the behavior of the mass of the meson in the bosonic term of the Lagrangian (\ref{L}), which approaches zero as $\lambda\to 1$.

\begin{figure}[t]
\centerline{\includegraphics[height=14em]{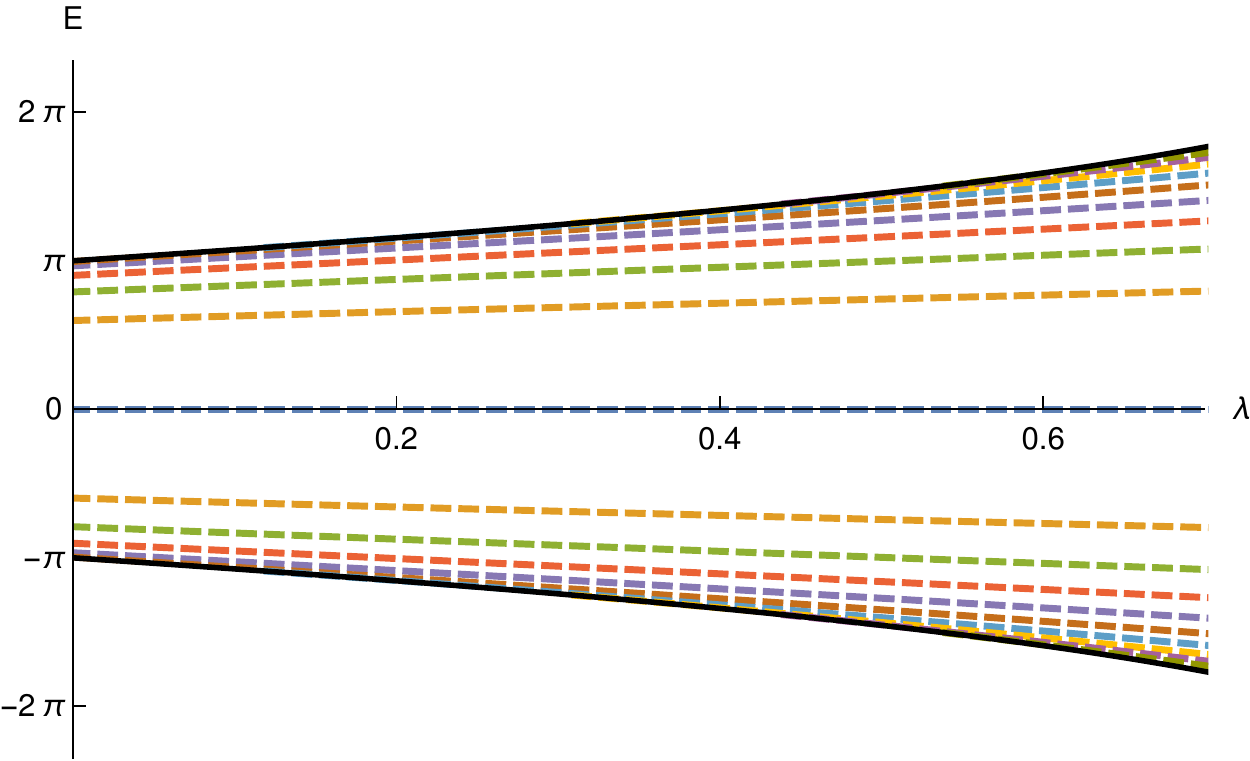}}
\caption{The fermionic bound state energy spectrum as a function of $\lambda$ for the model II. The solid curves identify the threshold energies.}\label{fig6}
\end{figure}

Once we have chosen the  solution (\ref{solmodel2}) as background field, we can search for the potential $U_-(x)$ to write
\begin{eqnarray}\label{Upmmodel2}
U_{-}\!\!&=&\!\!16 \left(\text{sc}^{-1}\!\left(\!\sqrt{\frac{1+\lambda}{1-\lambda}}\tanh\left(\frac{1}{2}\sqrt{1-\lambda ^2}x\right)\!,\!\lambda\!\right)\!\right)^2\\
\nonumber&~&-\frac{2\!\left(\!1-\!\lambda ^2\right)\!\text{nd}\!\left(\!\text{sc}^{-1}\!\left(\!\sqrt{\!\frac{1+\lambda}{1-\lambda}}\!\tanh\!\left(\frac{1}{2}\sqrt{1\!-\!\lambda^2}x\right)\!,\!\lambda\!\right)\!,\!\lambda \right)}{\cosh\!\left(\!\sqrt{\!1-\!\lambda ^2}x\!\right)\!-\!\lambda},
\end{eqnarray}
where $\text{nd}(x,\lambda)=1/\text{dn}(x,\lambda)$. This potential is depicted in Fig.~\ref{fig5}. Unlike the previous model, we now have a system in which the number of bound states increases as $\lambda$ also increases; so, we are ``capturing" or including bound states as $\lambda$ increases. This is illustrated in Fig.~\ref{fig6}. In particular, when $\lambda = 1$, where the fermionic potential becomes
\begin{equation}\label{UpmL1}
\left. U_{\pm}\right|_{\lambda=1}=16 \sinh ^{-1}(x)^2\pm \frac{4}{\sqrt{x^2+1}}
\end{equation}
we find an infinite tower of bound states.

\begin{figure}[t]
\centerline{\includegraphics[height=15em]{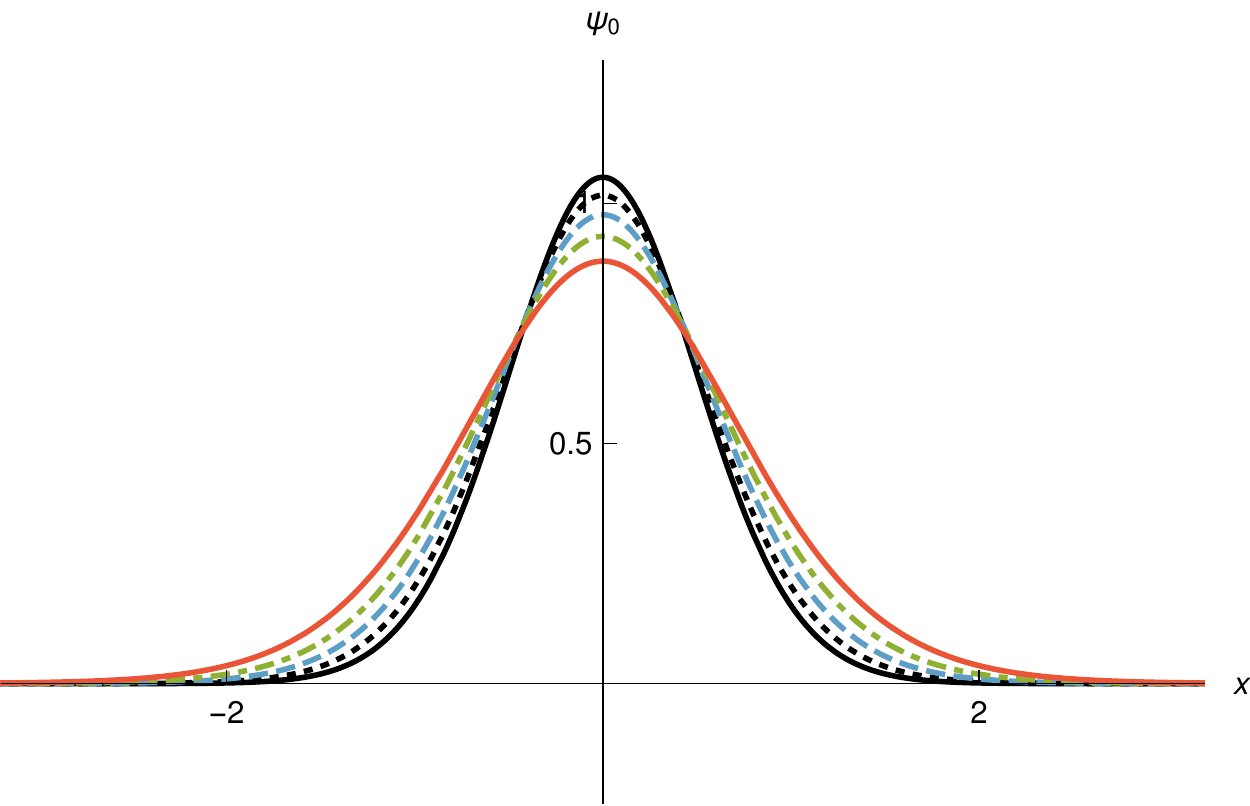}}
\caption{The normalized fermion zero mode derived from the solution \eqref{solmodel2} for $\lambda=0, 0.25, 0.5, 0.75,$ and $1$, depicted with solid (orange), dot-dashed, dashed, dotted and solid (black) curves, respectively, for the model II.}\label{fig7}
\end{figure}

Again, we can not calculate analytically the zero energy solution for the Dirac field for general $\lambda$, but we can still get some information about its behavior. In the neighborhood of $x = 0$, the scalar field behaves as $\phi(x\simeq 0)\simeq (1+\lambda) x+\mathcal{O}\left(x^2\right)$, so the wave function has the form $e^{-(1+\lambda) x^2+\mathcal{O}\left(x^3\right)}$. However, we should be careful when analyzing its asymptotic behavior because of the fact that the form of the solution in this regime is $\phi(x\to\infty)\simeq e^{-\sqrt{1-\lambda^2}x}+\phi_\infty$ with $\phi_\infty= 2 \text{sc}^{-1}\left(\sqrt{\frac{1+\lambda}{1-\lambda}},\lambda \right)$, which does not allow us to study the particular case $\lambda=1$. However, a direct analysis in the vacuumless solution shows that the asymptotic behavior of the scalar field is in the form $\phi_{\lambda=1}(x\to\infty)\simeq -2\ln x$. Thus, the ground state wave function decays as $e^{-2\phi_\infty x}$ for $\lambda\neq1$ and it decays as $e^{4x}x^{-4x}$ for $\lambda=1$. We can integrate the vacuumless solution in order to find the exact form of the ground state at $\lambda=1$, which is 
$$\psi(x,t)=c_+\left(
\begin{array}{c}
 e^{-4\left(x \sinh ^{-1}(x)-\sqrt{x^2+1}\right)}\\
0\\
\end{array}
\right).$$
In  Fig. \ref{fig7} the normalized zero mode is displayed for some values of $\lambda$. Here we note that it remains well behaved for all possible values of $\lambda$, including $\lambda=1$, and although there is an increase in its height, it behaves nicely in the full interval
$\lambda\in[0,1]$. This behavior is different from the one shown in the previous model, since there the zero mode shrinks to a narrower and narrower region around its core $x\approx0$ as $\lambda$ increases toward unity. 

\begin{figure}[t]
\centerline{\includegraphics[height=14em]{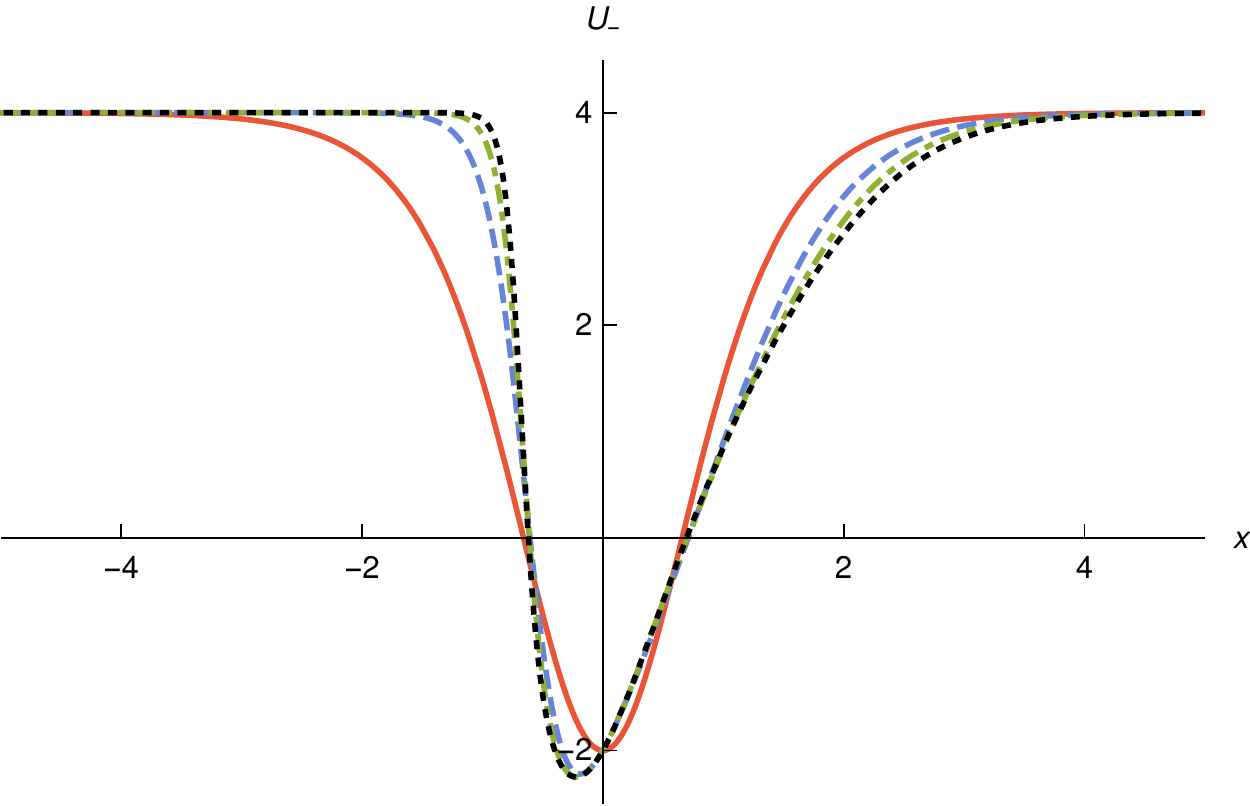}}
\caption{The $U_{-}$ potential of the model III, depicted for $p=1,3,5,$ and $7$ with solid, dashed, dot-dashed and dotted curves respectively.}\label{fig8}
\end{figure}
\begin{figure}[t]
\centerline{\includegraphics[height=14em]{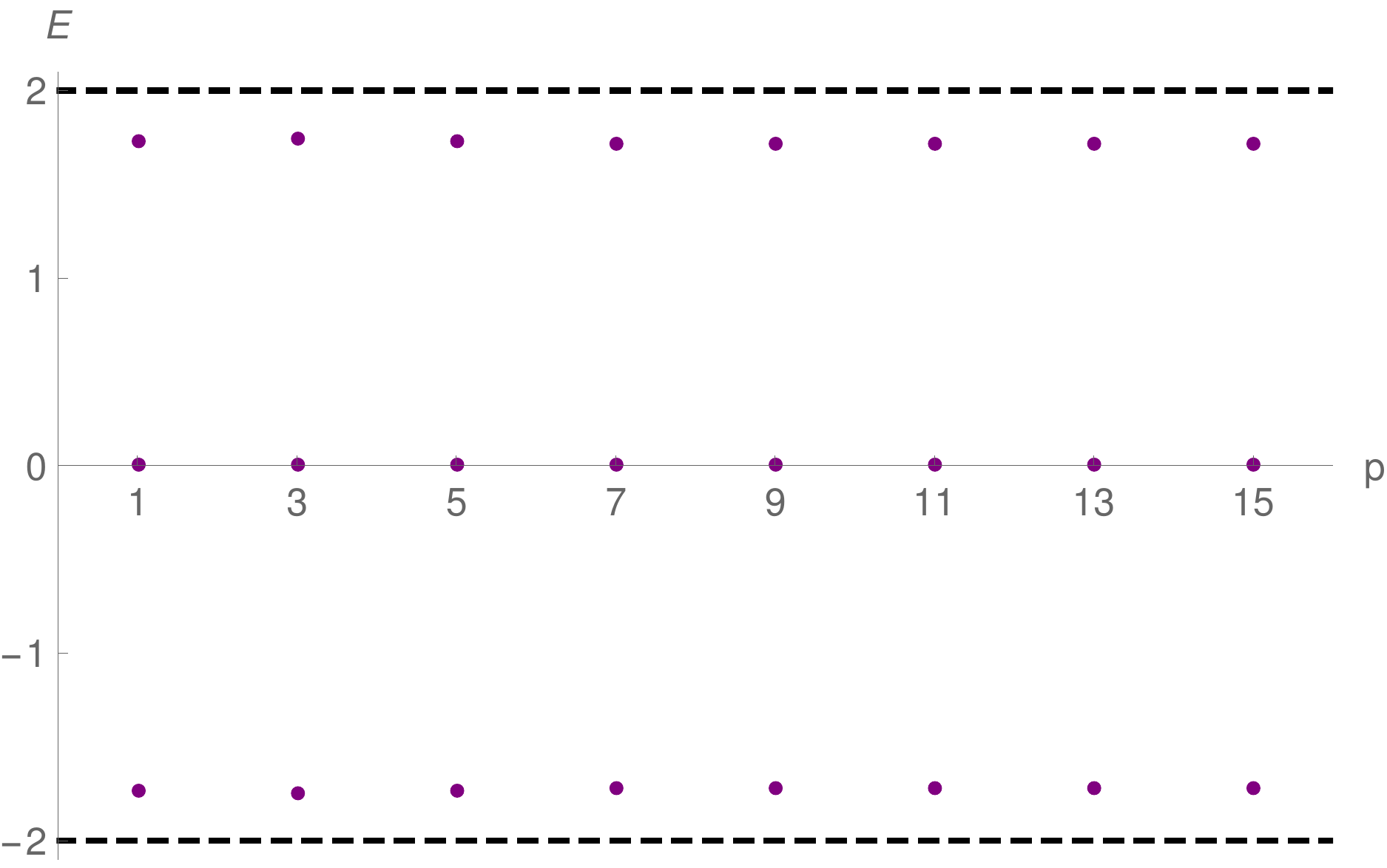}}
\caption{The fermion bound energy spectrum as a function of $p$ considering the model III. The dashed lines show the threshold energies.}\label{fig9}
\end{figure}

\subsection{Model III}

We now study how asymmetries within the scalar potential can affect the behavior of the fermionic bound states. 
We perform the numerical analysis of the model \eqref{asy}, presented in \cite{bmmashyb}. This model presents a topological sector between $\phi = 1 $ and $\phi = -1 $, where the masses of the mesons are given by $4$ and by $4p^2$, respectively. Note that as the scalar field asymptotically approaches $\phi(x\to\pm\infty)\to\pm1$, the height and width of the well remains almost the same for all $p$, unlike what happens with the stability potential for the bosonic field. Thus, the difference between the masses of the mesons in the scalar potential generated by the variations of the parameter $p$  implies only internal asymmetries in the fermion potentials, as depicted in Fig.~\ref{fig8}. Note that as the parameter $p$ increases, the fermionic potential presents higher asymmetry, breaking the reflection symmetry. As one can see, for negative values of $x$ the potential reaches the maximum value faster as $p$ increases. This is in contrast with the behavior for positive $x$, which is smoother.

\begin{figure}[t]
\centerline{\includegraphics[height=16em]{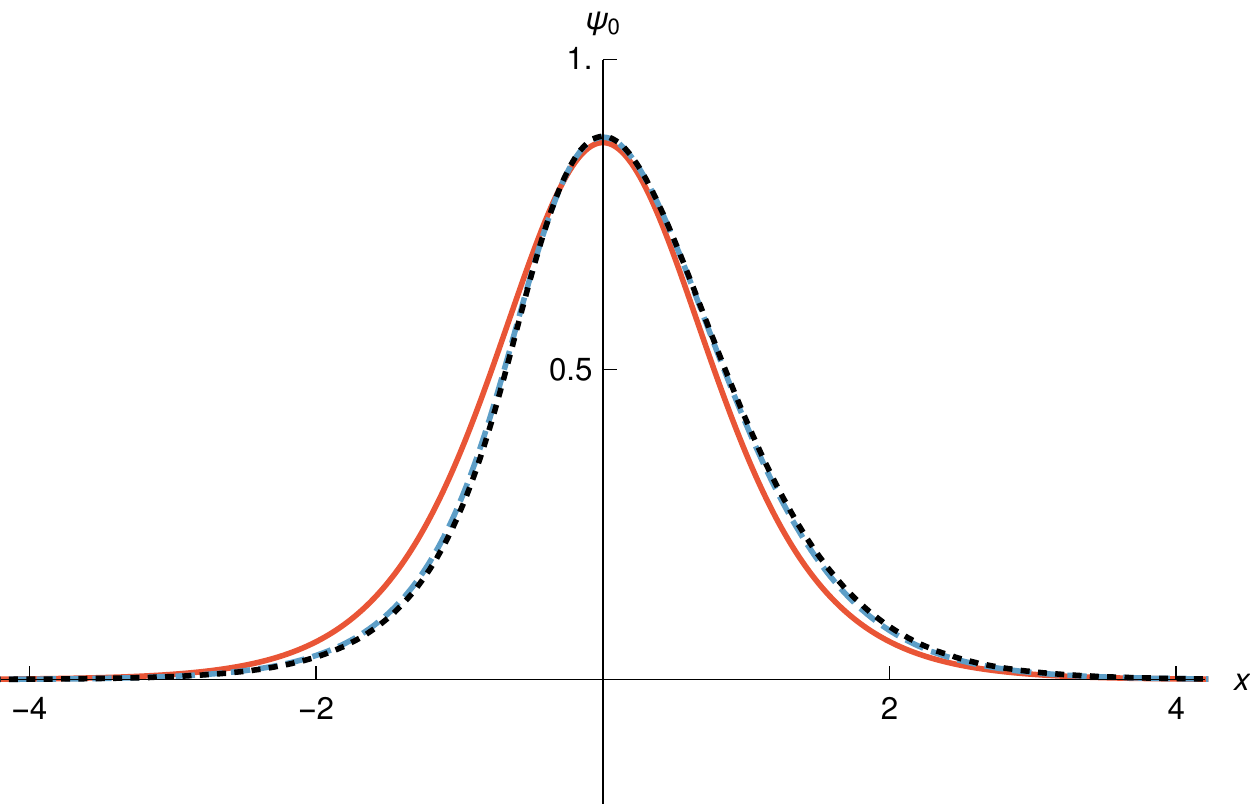}}
\caption{The normalized fermion zero mode derived from the model III, displayed for $p=1, 3$ and $5$ with solid (orange), dashed (blue) and dotted (black) curves, respectively.}\label{fig10}
\end{figure}
\begin{figure}[t]
\centerline{\includegraphics[height=16em]{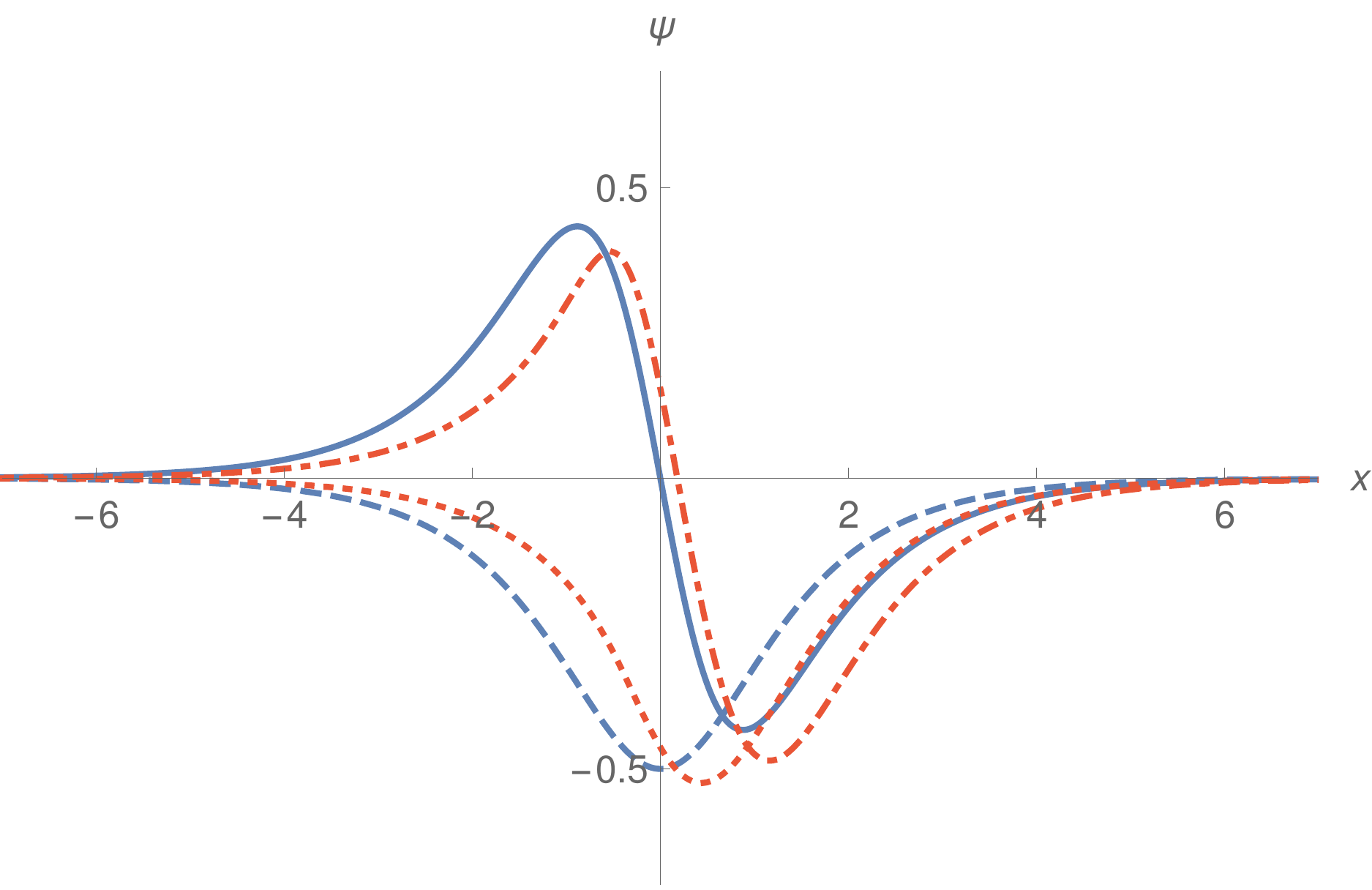}}
\caption{The $\psi^{(+)}$ and $\psi^{(-)}$ components of the massive bound states of the model III, depicted for $p=1$ with solid and dashed blue curves, and for $p=3$ with dot-dashed and dotted orange curves, respectively.}\label{fig11}
\end{figure}

In Fig.~\ref{fig9} we show the fermion bound energy spectrum for several values of the parameter $p$. As one can see, the energy is not much sensitive to the value of $p$, although it is not completely independent of $p$. In this model there are exactly three fermion bound energy states.
The normalized fermionic zero mode is depicted in Fig.~\ref{fig10}, where we observe that the shape varies only slightly as $p$ increases. It happens because asymptotically we have $\phi(x\simeq\infty)\simeq 1-e^{-2x}$ and 
$\phi(x\simeq-\infty)\simeq -1+e^{2p x}$. It implies that in the regime  $x\simeq\infty$ the ground state wavefunction decays as $\simeq e^{-2 x-e^{-2 x}}$ and in the limit $x\simeq-\infty$ it falls off as $\simeq e^{2 x+\frac{1}{p}e^{2 p x}}$. It means that for $ p> 1 $, the emerging nonlinearities due to the variations of this parameter are stronger for $ x <0 $. Moreover, the asymmetry of the fermionic ground state evolves as a function of $ p $ more slowly than the bosonic zero mode asymmetry, presented in \cite{bmmashyb}. This is due to the fact that the field nonlinearities appears in the exponent of the exponential, and thus the responses given in the curve format are less expressive. Therefore, in this model we notice that the fermionic zero mode responds asymmetrically to the parity-symmetry breaking.

\begin{figure}[h!]
\centerline{\includegraphics[height=16em]{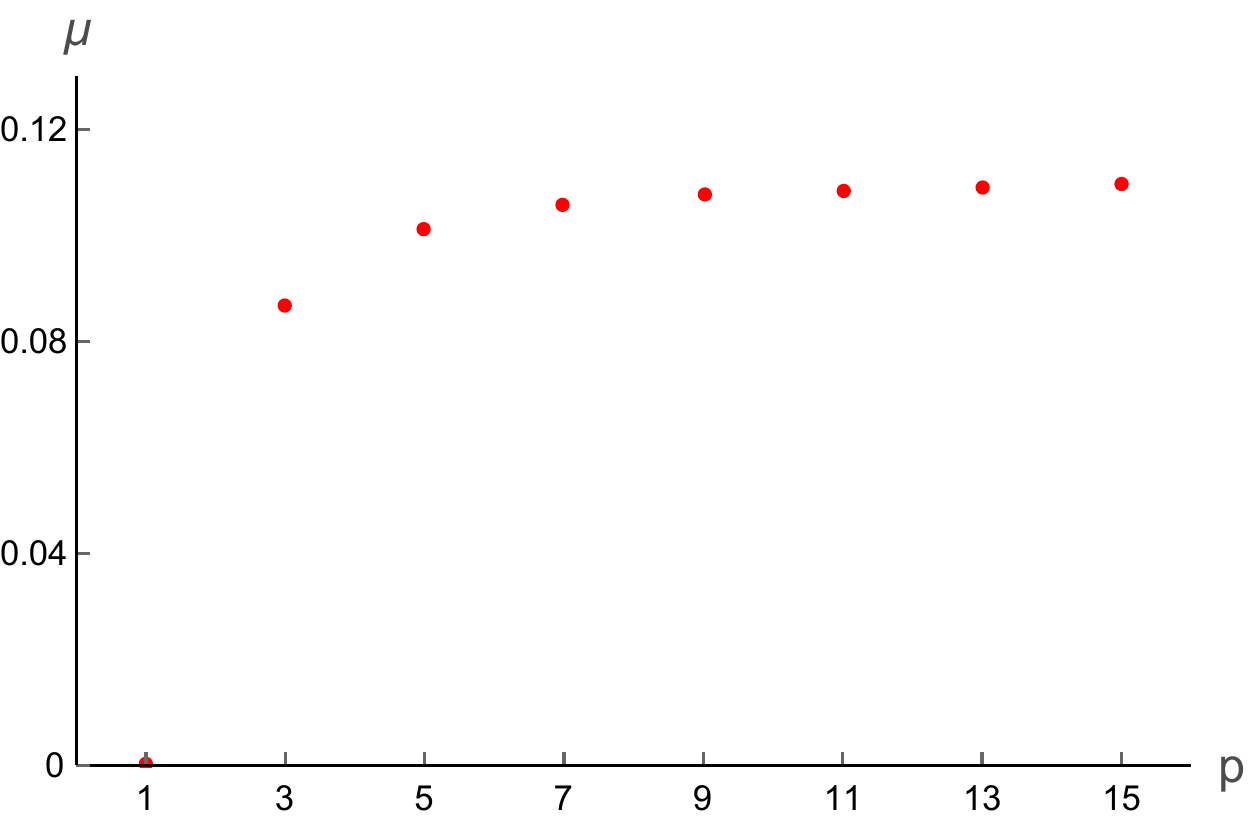}}
\caption{The mean value $\mu$ which measures the spatial asymmetry of the normalized zero mode, displayed for several values of $p$.}\label{fig12}
\end{figure}

Besides that, we also show the fermion massive bound states for the cases $p=1$ and $p=3$ in Fig.~\ref{fig11}. There one notices that the components $\psi^{(+)}$ and $\psi^{(-)}$ respect the parity symmetry for $p=1$, but this is not true anymore for the case $p=3$, as expected.

We can use the results depicted in Fig.~\ref{fig10} to quantify the asymmetry of the normalized zero mode via the mean value
\be
\mu={\int_{-\infty}^\infty dx\, x\, \psi_0^2}.
\ee
The results are displayed in Fig.~\ref{fig12}, where one sees no asymmetry for $p=1$, although it appears for $p=3,5,...$, varying smoothly as $p$ increases to larger values.

\section{Comments and conclusions}
\label{end}

In this work we studied the behavior of the fermion field in the background of three kinklike structures that respond with distinct geometric conformations. The three bosonic structures arise from models described by a single real scalar field recently investigated with distinct motivations, but here we use them to see how the fermion bound energies and states behave in each case. The two first models are controlled by a real parameter, $\lambda$, which highlight fascinating characteristics of the models. The third model is different and is controlled by an odd integer parameter, $p$, which induces the parity-symmetry breaking, due to the asymmetric form of the bosonic potential.   

The model I has the peculiarity of describing a background potential for the fermion field, which deepens and narrows as $\lambda$ increases towards unity, in a way such that the presence of fermion bound states diminishes with the increasing of the parameter. The model II has a distinct behavior, and the background potential is now capable of adding new fermion bound states as the parameter $\lambda$ increases in the interval $[0,1]$. We then see that for $\lambda$ increasing from zero to unity, while in the model I the number of fermion bound states diminishes, it increases unlimitedly in the model II.

While the models I and II obey parity symmetry, the model III engenders another behavior, with is also of current interest. It is controlled by an odd integer $p=1,3,5,...$, which is capable of inducing the parity-symmetry breaking. The calculations here are more intricate, but we have been able to show that the asymmetry present in the bosonic background is also induced in the potential of the fermion field, making the zero mode and the other bound states asymmetric. The asymmetry appears in the background potential and in the fermion bound states and may be considered for practical use, when one deals with asymmetric background structures; see, e.g., Ref.~\cite{As1}, where the asymmetry of the localized structure has played crucial role for the understanding of the kink-antikink collisions in the $\phi^6$ model, and also Ref.~\cite{As2} for the case of asymmetric structures in magnetic materials. 

\acknowledgements{D.B. and D.M. thank the Brazilian agencies CAPES and CNPq for financial support, and A.M. thanks PNPD/CAPES for the financial support.}


\end{document}